\documentclass[seceq,preprint]{ptptex}

\usepackage{graphicx}

\newcommand{\bzeta}{\mbox{\boldmath{$\zeta$}}}
\newcommand{\bgamma}{\mbox{\boldmath{$\gamma$}}}
\newcommand{\balpha}{\mbox{\boldmath{$\alpha$}}}
\newcommand{\bu}{\mbox{\boldmath{$u$}}}
\newcommand{\bq}{\mbox{\boldmath{$q$}}}
\newcommand{\bov}{\mbox{\boldmath{$v$}}}
\newcommand{\by}{\mbox{\boldmath{$y$}}}
\newcommand{\bxi}{\mbox{\boldmath{$\xi$}}}
\newcommand{\boeta}{\mbox{\boldmath{$\eta$}}}

\newcommand{\bolm}[1]{\mbox{\boldmath{$#1$}}}

\preprintnumber[15cm]{
\baselineskip0.5cm
OCU-PHYS-283, 
AP-GR-50, 
YITP-07-77, 
OU-TAP-284, 
APCTP Pre2008-004}

\title{
Magnification Probability Distribution Functions 
 \\of Standard Candles in a Clumpy Universe%
}

\author{
Chul-Moon \textsc{Yoo},$^{1,2,3}$ Hideki \textsc{Ishihara},$^2$ 
Ken-ichi \textsc{Nakao}$^2$ \\and 
Hideyuki \textsc{Tagoshi}$^4$ 
}

\inst{\small
$^1$Yukawa Institute for Theoretical Physics, \\
Kyoto University
Kyoto 606-8502, Japan\\
\medskip
$^{2}$Department of Physics,~Graduate School of Science, \\
~Osaka City University Osaka 558-8585,~Japan\\
\medskip
$^{3}$Asia Pacific Center for Theoretical Physics, \\
Pohang University of Science and Technology, Pohang 790-784,~Korea\\
\medskip
$^{4}$Department of Earth and Space Science, Graduate School of Science,\\
Osaka University, Toyonaka 560-0043, Japan\\
}

\recdate{
July 16, 2008}

\abst{
Lensing effects on light rays from 
point light sources, such like Type Ia supernovae, are simulated in a clumpy universe model. 
In our universe model, it is assumed that all matter in the universe 
takes the form of randomly distributed objects each of 
which has finite size and is transparent for light rays. 
Monte-Carlo simulations are performed for several 
lens models, and we compute 
probability distribution functions of magnification. 
In the case of the lens models that have a smooth 
density profile or the same degree of density concentration 
as the spherical NFW (Navarro-Frenk-White) lens model at the center, 
the so-called gamma distributions fit well 
the magnification probability distribution functions 
if the size of lenses is sufficiently larger than the Einstein radius. 
In contrast, the gamma distributions do not fit 
the magnification probability distribution 
functions 
in the case of the SIS (Singular Isothermal Sphere) lens model. 
We find, by using the power law cusp model,
that the magnification probability distribution function 
is fitted well using the gamma distribution only 
when the slope of the central density 
profile is not very steep. These results suggest that we may obtain information
about the slope of the central density profiles of dark matter halo 
from the lensing effect of Type Ia supernovae.
}

\begin{document}

\maketitle

\section{Introduction}\label{sec:SNeintro}

Type Ia supernovae (SNe)  are very useful tools to investigate our
universe. For example, the distance-redshift relation obtained as a 
result of the
observations of these strongly suggests the present acceleration of the
cosmic volume expansion of our universe\cite{Riess:1998cb,Perlmutter:1998np,
Knop,Riess:2004nr}. 
To confirm this
indication, further projects, such as the ESSENCE project at
NOAO,\footnote{http://www.ctio.noao.edu/wproject/}
the Large Synoptic Survey
Telescope\footnote{http://www.lsst.org/lsst\_home.shtml}
and Supernova/Acceleration Probe (SNAP),\footnote{http://snap.lbl.gov} are
now active or planned. 

Current observational data points on the distance-redshift
plane are somewhat scattered, and this is an origin of the error in
estimating the cosmological parameters. 
One of the reasons for this
distance dispersion is gravitational lensing effects due to the mass 
inhomogeneities in our universe. 
Although the dispersion due to gravitational lensing is 
relatively smaller than other effects at low redshift, 
it may become more prominent and comparable to the 
intrinsic dispersion at high redshift $z\gtrsim 1.2$
\cite{Sarkar:2007sp,Holz:2004xx}. 
In this sense, 
the mass
inhomogeneities are obstacles to the determination of the cosmological
parameters through the observation of Type Ia SNe. 
On the other hand, the mass inhomogeneities contain rich information about the physical 
process of the evolution of our universe. 
Thus, the mass inhomogeneities themselves are very significant subjects in cosmology.

We can understand the property of the mass inhomogeneities 
by comparing the observational data about type Ia SNe with theoretical
predictions obtained through the investigations of the gravitational lensing
effects on the light rays from them. 
This is just the purpose of this paper. 

There are many works about 
gravitational lensing related to SNe: effects on SNe observations
\cite{1991ApJ...374...83R,
1997ApJ...475L..81W,
Frieman:1996xk,
1998ApJ...506L...1H,
2000ApJ...532..679P,
2000MNRAS.318..195B,
2000A&A...354..767V,
Holz:2004xx,
2006PhRvL..96b1301C,
Martel:2007fh}, 
their availability for investigation of our universe 
\cite{1999ApJ...519L...1M,
1999MNRAS.305..746M,
1999A&A...351L..10S,
2000ApJ...534...29H,
2001ApJ...559...53M,
2001ApJ...556L..71H,
2007arXiv0711.0743L}, 
the extraction of the evidence for lensing effects from 
observational data 
\cite{2004MNRAS.351.1387W,
2005MNRAS.358..101M,
Wang:2004ax,
2007JCAP...06....2J} 
and others
\cite{1998PhRvD..58f3501H,
2000PhLB..486..249G,
Wang:1999bz,
2002MNRAS.335.1061S,
Wang:2002qc,
Gunnarsson:2003fy,
Wang:2003gz,
2006astro.ph..1683M,
2007PhRvL..98g1302M}. 

In this paper, we particularly focus on the effects of small-scale structures. 
The smaller scale inhomogeneity may affect 
the apparent magnitude of sources 
owing to gravitational focusing. 
The mass scale of the inhomogeneities, which is considered in this paper, is 
between $10^{-2}M_\odot$ and $10^{10}M_\odot$. 
There are several works related to this subject
\cite{1998PhRvD..58f3501H,1999ApJ...519L...1M,2007PhRvL..98g1302M,2001ApJ...559...53M,1999A&A...351L..10S}. 
In these works, compact lens objects were mainly studied, whereas,
in this paper, we consider the extended lens objects. 
Even in the case of the extended lens objects, 
the strong lensing effects that cause multiple images are 
of considerable importance if the density profiles 
of the objects are significantly steep at the center. 

We  consider  the flat $\Lambda$CDM model 
with $\Omega_{\Lambda 0}=0.71, \Omega_{\rm m0}=0.29$ as the starting point. 
In the cold dark matter cosmology, 
the first objects to form are of subgalactic size. 
Then, larger structures form through tidal interaction and mergers of smaller 
objects. 
Therefore, 
the dark matter may form numerous clumps. 
For simplicity, we assume that all matter in the universe takes 
the form of randomly distributed objects, each of
which has finite size and is transparent for light rays.
We investigate the magnification effect 
due to the gravitational 
lensing on light rays from the point sources such as Type Ia supernovae.
The light rays from distant supernova events 
suffer multiple gravitational lensing effects from 
the clumps in the universe. 
The propagation of the light rays is thus stochastic owing to 
the randomness of the distribution of the clumps. 
In this paper, we focus on 
magnification probability distribution function (MPDF). 
We introduce a simple method to simulate multiple 
gravitational lensing effects in the clumpy universe model, 
which is based on our previous study \cite{Yoo:2006gn}.
We perform Monte-Carlo simulations for several lens models, 
and we investigate the dependence of the MPDF on lens models of 
dark matter clumps.

Throughout this paper, 
it is assumed that light sources are pointwise. 
This assumption is valid only when the linear extent of 
a source is much smaller than the Einstein radius of each 
lens object. 
The linear extent of the SNe is of the order of $\sim10^{15}$ cm, 
which is the typical radius of 
SNe at the peak brightness. 
The Einstein radius of a lens object of mass $M_L$ is of the order 
$\sqrt{GM_LD/c^2}$, where 
$D$ is the distance to supernovae, $G$ and $c$ are 
the gravitational constant and speed of light, respectively. 
Therefore, we assume that 
$M_L$ satisfies $\sqrt{GM_LD/c^2}\gg10^{15}{\rm cm}$. 
Because in the case of high-redshift supernova events, $D\sim 1{\rm Gpc}$, 
this assumption leads 
\begin{equation}
M_L\gg10^{-2}M_\odot.
\end{equation}

Our method is complementary to the methods using the N-body simulation. 
In the N-body simulation, the nonlinear time evolution and the spatial correlation 
of the dark matter distribution can be studied. 
Very large scale N-body simulations are possible now. 
Nevertheless, the questions that can be definitively 
answered with N-body simulations 
are still limited by finite resolution in mass and in distance 
\cite{BinneyTremaine}. 
Since the small-scale structure of the dark matter is 
essential for the
gravitational
lensing effects, 
the simulations using analytic density profiles 
are important to confirm the results from the N-body simulation, 
and to investigate new effects that are not found in the N-body simulations.

This paper is organized as follows. 
In \S\ref{sec:rev}, we briefly review the gravitational 
lensing. 
\S\ref{sec:cluni} is devoted to describe the clumpy 
universe model. 
In \S\ref{sec:mag}, we show a definition of 
magnification and the relation between the distance and 
the magnification. 
In \S\ref{sec:nucalre}, numerical calculation and our results are 
illustrated. 
Finally, \S\ref{sec:consum} is devoted to the conclusions and summary.
Throughout this paper, we use the unit of $G=c=1$. 

\section{Gravitational lensing in a clumpy universe model}

\subsection{Lens equation} \label{sec:rev}

\begin{figure}[htbp]
 \begin{center}
  \includegraphics[scale=0.8]{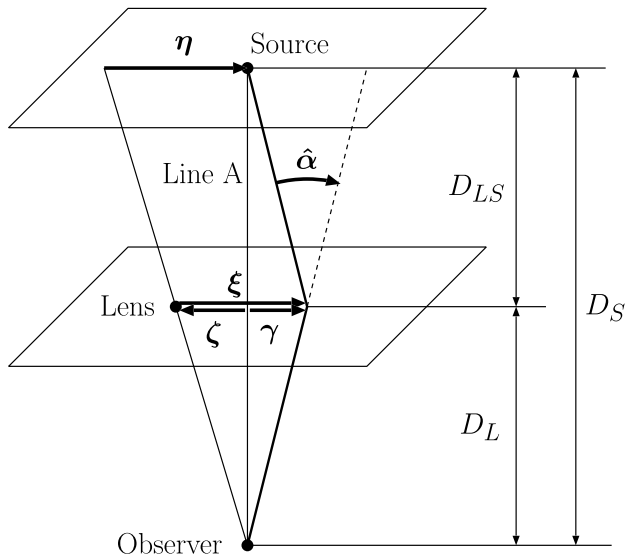}
  \caption{Geometry of gravitational lensing due to a point-mass lens.
  The impact vector $\bxi$ represents the relative position of 
  a light ray on the lens plane to the lens position, and 
  $\hat{\balpha}$ is a vector whose norm is equal to 
  the bending angle.} 
  \label{fig:lens}
 \end{center}
\end{figure}

Before describing our settings and calculation method, 
let us briefly review gravitational lensing. 
In this paper, we focus on the situation in which 
thin lens approximation\cite{SEF} 
is valid. 
The thin lens models are characterized by 
the surface mass density $\Sigma(\bxi)$, 
where $\bxi$ is the impact vector (see Fig. \ref{fig:lens}). 
By using the surface mass density $\Sigma(\bxi)$,  
the bending angle vector $\hat{\balpha}(\bxi)$ is given
as~\cite{SEF}
\begin{equation}
\hat{\bolm\alpha}(\bolm\xi)
=4\int\frac{(\bolm\xi-\bolm\xi')
\Sigma(\bolm\xi')}{|\bolm\xi-\bolm\xi'|^2}d^2\xi'. \label{eq:defangle}
\end{equation} 
%
%
For convenience, we consider the ``straight" line $A$ 
from a source to the observer and define the intersection point 
of $A$ with the lens plane as the origin of the lens position 
$\bzeta$ and the ray position $\bgamma$. 
In the geometrically thin lens approximation, we can regard 
the vectors $\bxi$, $\bzeta$, and $\bgamma$ on the lens plane~\cite{SEF}. 

We define $D_S$, $D_L$, and $D_{LS}$ as the angular diameter 
distances from the observer to
the source, from the observer to the lens, and from the lens to the
source, respectively.
Using simple trigonometry~(see Fig. \ref{fig:lens}),
we find the relation between the source position $\boeta$ and
$\bxi$, 
\begin{equation}
 \boeta
 = \frac{D_S}{D_L} \bxi - D_{LS} \hat{\balpha}(\bxi).
 \label{eq:lens}
\end{equation}
From Eq.~(\ref{eq:lens}), we have 
\begin{equation}
 0
 = \frac{D_S}{D_L} \bgamma
   - D_{LS} \hat{\balpha}(\bgamma - \bzeta)
\label{eq:lens2}
\end{equation}
by using $\bgamma$ instead of $\bxi$ and $\boeta$. 

\subsection{Clumpy universe model}\label{sec:cluni}

The clumpy universe model 
in which we calculate the lensing effects is 
basically the same as that in Ref. \citen{Yoo:2006gn}. 
In this clumpy universe model, 
it is assumed that all matter 
takes the form of randomly distributed objects
and its global property 
is well described using the Friedmann-Lema\^itre (FL) 
universe whose metric is given as 
\begin{equation}
 ds^2
 = - dt^2
   + a^2(t) \left(\frac{dr^2}{1 + K r^2} + r^2 d\Omega^2 \right), 
\label{eq:homet}
\end{equation}
where $K=1$, $0$, and $-1$, and $d\Omega^2$ is the round metric. 
Hereafter, we will refer to this FL universe as 
``background universe". 

\begin{figure}[htbp]
\begin{center}
\includegraphics[scale=0.7]{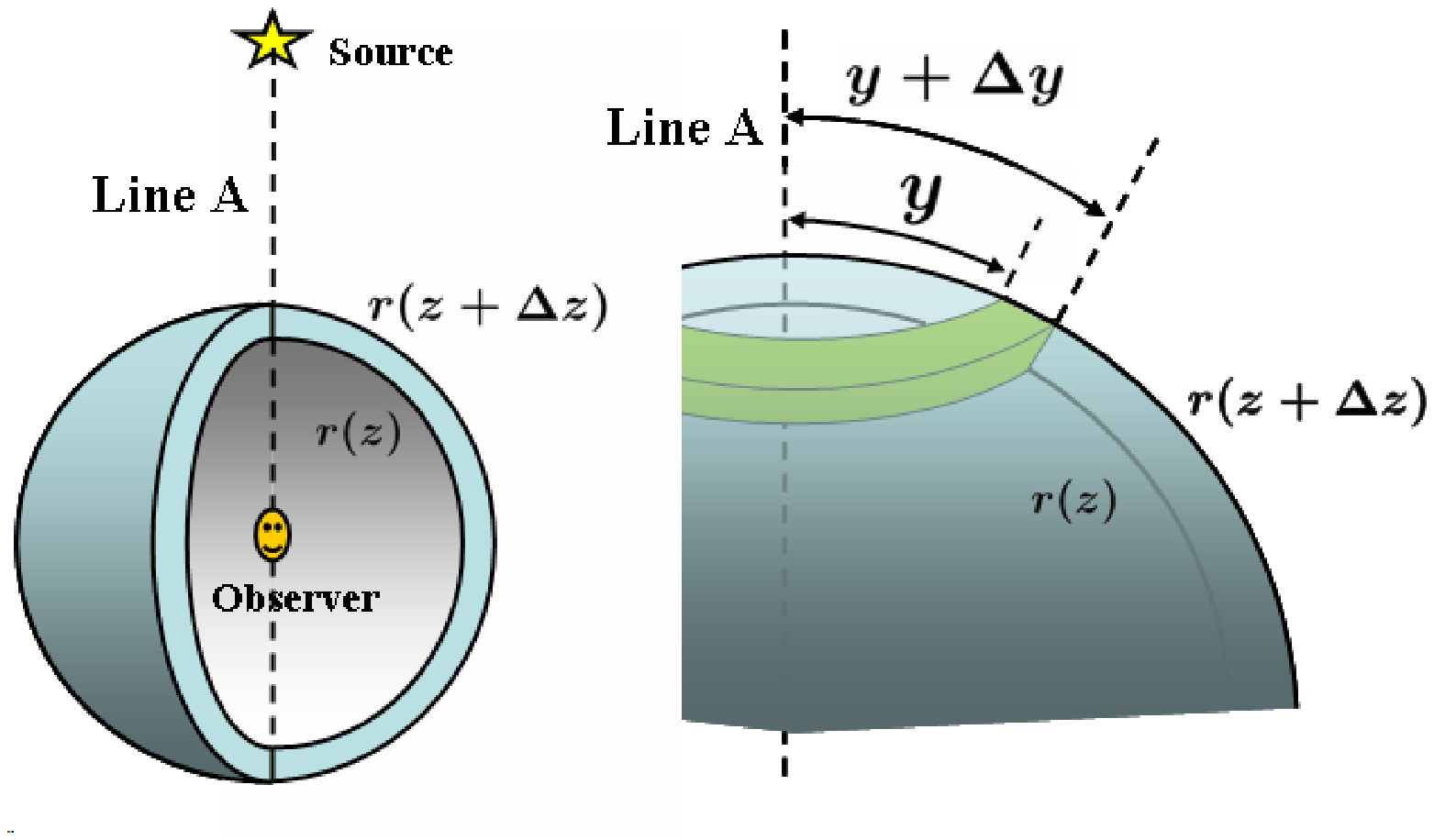}
\caption{Figures for \S\ref{sec:cluni}.
}
\label{fig:lendis}
\end{center}
\end{figure}

We assume that the comoving number density 
$\rho_{\rm n}$ of the lenses in the clumpy universe is given as 
\begin{equation}
 \rho_{\rm n}
 = \frac{a^3 \rho}{M_L}
 = \frac{3 \Omega_{\rm m0} H_0^2}{8 \pi M_L}, 
\end{equation}
where $\rho$, $\Omega_{\rm m0}$, and $H_0$ are the average mass density, 
the present values of the total density parameter, and 
the Hubble parameter, respectively. 
We consider a past light cone of the observer at $r=0$ 
in the background universe, which is parametrized by the 
redshift $z$ of its null geodesic generator. 
The comoving volume $\Delta V$ 
of a spherical shell (see Fig. \ref{fig:lendis})
bounded by $r=r(z)$ and  
$r(z+\Delta z)$ on this light cone is given as 
\begin{equation}
 \Delta V = \frac{4 \pi r^2 }{\sqrt{1+Kr^2}}\frac{dr}{dz} \Delta z. 
\end{equation}
Therefore, the number of lenses in this shell is given as  
\begin{equation}
 \Delta N
 = \rho_{\rm n} \Delta V
 = \frac{3 \Omega_{\rm m0} H_0^2r^2}{2 M_L\sqrt{1+Kr^2}}  \frac{dr}{dz} \Delta z. 
 \label{eq:dN}
\end{equation}
We assume $dr/dz$ is the same as the background value, 
\begin{equation}
 \frac{dr}{dz}
 = \frac{1}{H_0}
   \sqrt{\frac{1 + H_0^2 \Omega_{K0} (1 + z)^2 D_{FL}^2(z)}
              {\Omega_{\rm m0}(1 + z)^3 - \Omega_{K0} (1 + z)^2
               + \Omega_{\Lambda0}}
        },
\label{eq:dR}
\end{equation}
where $\Omega_{\rm \Lambda0}$ and $D_{FL}(z)$ are the 
normalized cosmological constant and 
the angular diameter distance 
from the source of the redshift $z$ to the observer in the 
background universe, respectively, 
and $\Omega_{\rm K0}=\Omega_{\rm m0}+\Omega_{\Lambda0}-1$.

Let us consider a lens plane at the redshift $z$. 
We define $\by$ as 
\begin{equation}
\by:=\frac{\bzeta}{\xi_0}, 
\end{equation}
where $\xi_0$ is 
the Einstein radius 
given as 
\begin{equation}
 \xi_0 = \sqrt{4 M_L \frac{D_L D_{LS}}{D_S}}. \label{eq:re}
\end{equation}
$\by$ represents 
the lens position normalized with the Einstein radius. 
In addition, we write the absolute values of $\by$ and $\bzeta$ as 
$y$ and $\zeta$, respectively. 
The average number of lenses in the region
$[y,y+\Delta y)$ within the domain $[z,z+\Delta z)$(see Fig. \ref{fig:lendis}) 
is given as 
\begin{equation}
\frac{2\pi \xi_0^2 y \Delta y}{4 \pi a^2 r^2} \Delta N, 
 \label{eq:p1}
\end{equation}
where we have assumed $\zeta\ll r$. 
This value is equivalent to the probability 
$p(y,z)\Delta y$ for a light ray 
to receive lensing effects with impact parameter $[y,y+\Delta y)$ in 
the domain $[z,z+\Delta z)$. 
Substituting Eqs.~(\ref{eq:dN}) and (\ref{eq:dR}) into 
Eq. (\ref{eq:p1}) and 
using Eq. (\ref{eq:re}), we have
\begin{equation}
 p(y,z) \Delta y
 = \frac{3 H_0 \Omega_{\rm m0} (1 + z)^2}{\sqrt{\Omega_{\rm m0}(1 + z)^3 - \Omega_{K0} (1 + z)^2
               + \Omega_{\Lambda0}}}
 \frac{D_{L} D_{LS}}{D_{S}}
 y \Delta y \Delta z.
 \label{eq:p3}
\end{equation}
It is seen from Eq. (\ref{eq:p3}) that this probability is specified 
using the 
model of the background universe.

\subsection{Magnification and distance}\label{sec:mag}

In this subsection, we discuss the definition of 
magnification due to lensing and the definition of distance
in the clumpy universe model. 

First, we consider a point light source that is observed as a 
single image in a clumpy universe. 
Magnification $\mu$ 
for a source is defined as
\begin{equation}
\mu=\frac{S}{S_0}, 
\label{eq:defmag}
\end{equation} 
where $S$ is the flux actually observed, whereas $S_0$ is
the fictitious flux that would be emitted by 
the source with the same luminosity and redshift but 
without lensing effects. 
If the light ray bundle passes through a spatial section with 
an area $\Delta A$ orthogonal 
to the ray bundle, 
using the conservation law of photon number\cite{SEF}, 
we find 
\begin{equation}
\mu=\frac{S}{S_0}=\frac{\Delta A_0}{\Delta A}, 
\end{equation}
where $\Delta A_0$ is a sectional area 
orthogonal to the ray direction in the 
fictitious propagation process.

The corrected luminosity distance is defined as 
\begin{equation}
D'=\left(\frac{\Delta A}{\Delta\Omega_S}\right)^{1/2}, 
\end{equation}
where $\Delta\Omega_S$ is the solid angle that subtends a light ray bundle 
from the point source. 
The relation between the angular diameter distance $D$ and the 
corrected luminosity distance $D'$ is given as 
$D'=(1+z_S)D$, 
where $z_S$ is the source redshift\cite{SEF}. 
Let $D_0$ denote the angular diameter distance 
between the source and observer in the case without any lensing effects. 
Since 
\begin{equation}
\left(\frac{\Delta A_0}{\Delta \Omega_S}\right)^{1/2}=(1+z_S)D_0, 
\end{equation}
we obtain 
\begin{equation}
D'
=\frac{\left(\frac{\Delta A}{\Delta\Omega_S}\right)^{1/2}}{\left(\frac{\Delta A_0}
{\Delta\Omega_S}\right)^{1/2}}(1+z_S)D_0
=(1+z_S)\frac{D_0}{\sqrt{\mu}}.
\end{equation}
Therefore, we can define the observed angular diameter distance $D_A$ as 
\begin{equation}
D_A=\frac{D_0}{\sqrt{\mu}}. 
\end{equation}

Next, let us consider the case of 
multiple images. 
The magnification $\mu_{(p)}$
of the $p$-th image of a source is defined as 
\begin{equation}
\mu_{(p)}=\frac{S_{(p)}}{S_0}, 
\end{equation} 
where $S_{(p)}$ is the flux of the $p$-th image. 
In this study, we assume that multiple images 
can not be distinguished from 
each other owing to the limitation of the resolution, 
and we can observe only the total flux 
of light rays. 
Then the observed magnification $\mu$ is given as 
\begin{equation}
\mu=\frac{\sum_pS_{(p)}}{S_0}=\sum_p\mu_{(p)}, 
\end{equation}
and we define 
the observed angular diameter distance as 
\begin{equation}
D_A=\frac{D_0}{\sqrt{\sum_p\mu_{(p)}}}. 
\end{equation}
In terms of the angular diameter distance, 
the magnification is written as 
\begin{equation}
\mu=\frac{D_0^2}{D_A^2}. 
\end{equation}

The angular diameter distance in gravitational lensing is 
discussed for many years\cite{Fukugita,Asada}. 
The angular diameter distance in the FL universe, $D_{FL}$, and the Dyer-Roeder distance 
\cite{1973ApJ...180L..31D,1969ApJ...155...89K}, $D_{DR}$, 
are commonly used. 
The Dyer-Roeder distance is the distance in the under dense region in 
the inhomogeneous matter distribution. 
It depends on the smoothness parameter $\alpha$, which is the ratio 
of the smoothly distributed matter density except for clumps to the 
mean energy density $\rho$ for all matter in the universe. 
When $\alpha=1$, the universe becomes the FL universe 
and $D_{DR}$ agrees with $D_{FL}$.
When $\alpha=0$, all matter takes the form of clumps, 
and in such a case, 
the Dyer-Roeder distance is the distance 
in an empty region in the clumpy universe. 

In our work, 
unlensed angular diameter distance $D_0$ should be appropriately 
chosen so as to be consistent with 
the situations of our present
interest as below
\cite{Seitz1994}. 
We are investigating the lensing effects of SNe,
whose linear extent is much smaller than 
the scales in which density inhomogeneities are linear or quasi-linear. 
Then, 
the matter distribution might be highly clumpy, 
and the under dense region might be 
almost empty.

In this situation, we expect that
the focusing effects the light ray receives are minimal,
and that the angular diameter distance agrees with the Dyer-Roeder
distance,
because the light rays do not receive any lensing effects
between the lens planes.
This is consistent with the numerical simulations 
by Holz \& Wald\cite{1998PhRvD..58f3501H} and Kozaki\cite{Kozaki}. 
Holz \& Wald\cite{1998PhRvD..58f3501H} discussed the
angular diameter distance in the universe that 
consists of many small balls.
Kozaki\cite{Kozaki} calculated 
the distance-redshift relations in the Swiss-cheese universe model, 
that is an exact solution of the Einstein equation.
In this paper, 
we adopt the Dyer-Roeder distance $D_{DR}$ with $\alpha=0$
as $D_0$ and for the distance between 
each lens plane which has positive definite 
surface mass density$\Sigma(\xi)$. 
Hereafter, we drop ``with $\alpha=0$",
and simply call this distance the ``Dyer-Roeder distance $D_{DR}$".

In Refs.\citen{Tomita,TAH}, it is shown that the observed distance
after the multiple lensing effects sometimes becomes longer than
the DR distance. 
In such cases, $\alpha$ becomes negative.
In our simulation, we do not treat such a case and 
use the Dyer-Roeder distance as the angular diameter distance. 
To understand such situations, 
we need further studies.

If inhomogeneities are so small that these are regarded as 
linear or quasi-linear perturbations in the homogeneous 
and isotropic universe, 
perturbative treatments seem to be better 
in gravitational lensing. 
In such perturbative treatments, 
the unlensed distance $D_0$ is the angular diameter distance in the 
FL universe $D_{FL}$. 
Light rays are perturbed by the lensing effects 
of the inhomogeneity. 
In this sense, the (de)magnification is often 
defined as $\tilde \mu:=D_{FL}^2/D_A^2$ in weak lensing analyses 
and ray shooting calculations\cite{Wambsganss,Premadi,Jain}. 
In their works, the surface mass density 
of each lens plane is not positive definite, namely, 
the under dense regions have negative mass density. 
The relation between $\mu$ and $\tilde \mu$
is given as 
\begin{equation}
\tilde \mu=\mu \frac{D_{FL}^2}{D_{DR}^2}.
\end{equation}

\section{Numerical calculation and results}\label{sec:nucalre}

\subsection{Numerical methods and the density profiles of lens models}

In this study, we use the numerical method 
proposed by Rauch\cite{1991ApJ...374...83R}. 
We compute lens mappings using
the multiple lens plane method~\cite{SEF}.
We ignore the lensing effects coming from 
clumps far from the ray.
We search for the position of images 
on the field of view using the Newton-Raphson method.
Summing up the magnifications of all images, 
we obtain the total magnification.
The concrete description about the method 
is in Ref.\citen{Yoo:2006gn} and we 
review it in Appendix \ref{sec:cal}. 

In the simulations, 
we assume that clumps in the universe are transparent to light rays. 
We also assume that clumps are axisymmetric, and 
the axes of symmetry of clumps are identical with the line of sight. 
Each of these clumps has the mass $M_L$ and the ``size" $R$.  
We consider the following three lens models:
\footnote{Numerical calculations have been performed 
also for homogeneous sphere:$\Sigma(\xi)
=\frac{3M_L}{2\pi R^2}\left(1-\xi^2/R^2\right)^{1/2}$ and 
power low tail model:$\Sigma(\xi)=\frac{M_L}{\pi R^2\left(1+\xi^2/R^2\right)^2}$. 
The results are qualitatively the 
same as those of the lens models (a) and (b). 
We do not show these results in this paper. }
\begin{itemize}
\item[(a)]{Homogeneous disk:}
\begin{equation}
\Sigma(\xi)=\left\{
\begin{array}{ccc}
\frac{M_L}{\pi R^2}&{\rm for}&\xi<R, \\
0&{\rm for}&\xi\geq R. 
\end{array}
\right.
\label{eq:homolen}
\end{equation}
\item[(b)]{Log cusp model:}
\begin{equation}
\Sigma(\xi)=\left\{
\begin{array}{ccc}
\frac{2M_L}{\pi R^2}\ln\frac{R}{\xi}&{\rm for}&\xi<R, \\
0&{\rm for}&\xi\geq R. 
\end{array}
\right. 
\end{equation}
\item[(c)]{$1/\xi$ cusp model (Singular Isothermal Sphere):}
\begin{equation}
\Sigma(\xi)=\left\{
\begin{array}{ccc}
\frac{M_L}{2\pi R^2}\frac{R}{\xi}&{\rm for}&\xi<R, \\
0&{\rm for}&\xi\geq R. 
\end{array}
\right. 
\end{equation}
\end{itemize}

The degree of the density concentration at the center 
of the log cusp model (b)
is equivalent to the spherical Navarro-Frenk-White (NFW)\cite{NFW} lens model 
in which 
the mass density, $\rho$, is inversely proportional to the 
radius.

The surface mass density of all our lens models can be 
written in the form, 
\begin{equation}
\Sigma(\xi)\propto \frac{M_L}{R^2}F\left(\frac{R}{\xi}\right), 
\label{eq:lensform}
\end{equation}
where $F(R/\xi)$ is a function of $R/\xi$, $\xi$ is 
the distance from the axis of symmetry on the 
lens plane, and $\Sigma(\xi)$ is the surface mass density.
We assume that the total mass of each lens object 
$M_L=2\pi\int^\infty_0\xi\Sigma(\xi)d\xi$ is finite. 
If we use Eq. (\ref{eq:lensform}) and use the calculation method 
written in Appendix \ref{sec:cal}, 
we find that the magnification distribution does not change 
if we scale the parameter in the manner $R\rightarrow CR$ 
and $M_L\rightarrow C^2M_L$, where 
$C$ is an arbitrary constant. 
This means that 
the magnification depends only on the parameter $R/\sqrt{M_L}$. 
A proof of this fact is given in Appendix \ref{sec:cal}.

\subsection{MPDFs}\label{sec:mpdfs}

\begin{figure}[htbp]
\begin{center}
\includegraphics[scale=1.13]{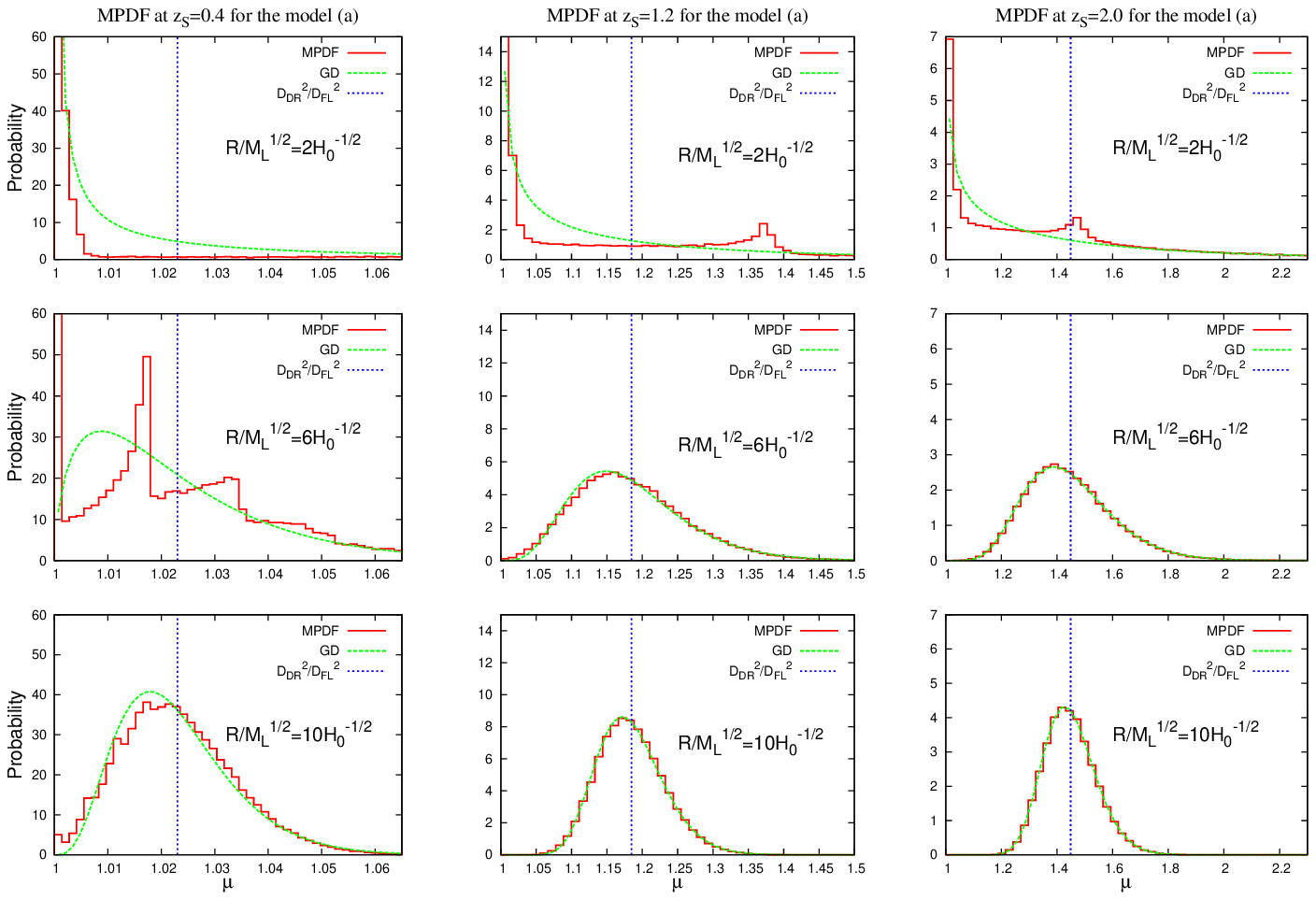}
\caption{MPDFs for the lens model (a) at 
$z_S=0.4$, $1.2$, and $2.0$ are shown on the left, center, and 
right lines, respectively. 
The smooth lines are the gamma distributions that 
fit the MPDFs.
}
\label{fig:Ldepdisk}
\end{center}
\end{figure}

\begin{figure}[htbp]
\begin{center}
\includegraphics[scale=1.13]{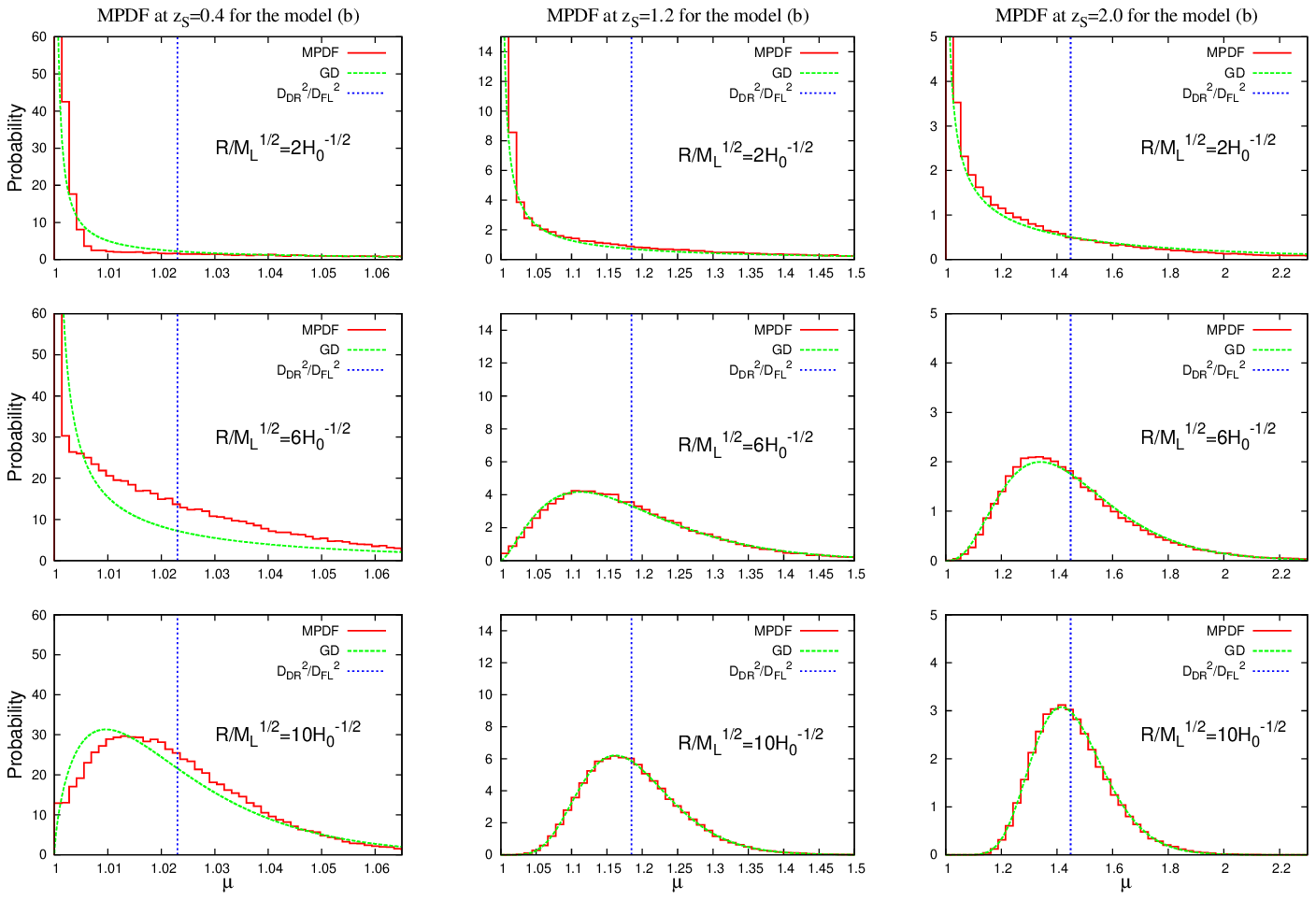}
\caption{Same as Fig. \ref{fig:Ldepdisk}, 
but for the lens model (b). 
}
\label{fig:Ldepnfw}
\end{center}
\end{figure}

\begin{figure}[htbp]
\begin{center}
\includegraphics[scale=1.13]{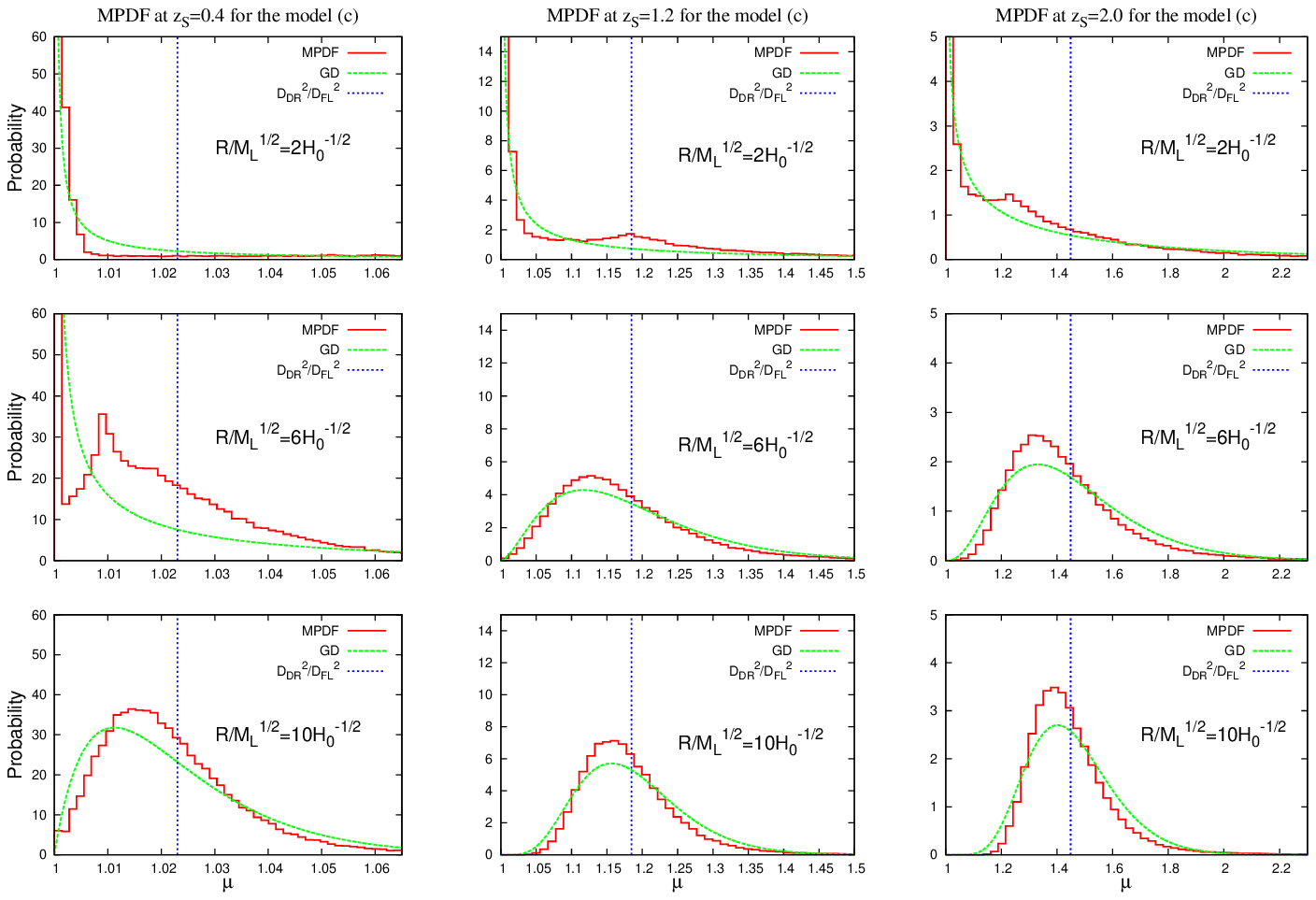}
\caption{Same as Fig. \ref{fig:Ldepdisk}, 
but for the lens model (c). 
}
\label{fig:Ldepsis}
\end{center}
\end{figure}

We first consider the cases in which $R/\sqrt{M_L}=$constant, 
and investigate the dependence of MPDFs on lens models and on the value of
$R/\sqrt{M_L}$. 
The magnification is stochastic owing to 
the randomness of the lens distribution.
We numerically generate 100,000 samples of magnification 
for each lens model characterized by the parameter $R/\sqrt{M_L}$. 
In our calculations, 
we use the same realization of the lens distribution for each model. 
The MPDFs given as results of these calculations are shown in 
Figs. \ref{fig:Ldepdisk}-\ref{fig:Ldepsis}. 

In Fig. \ref{fig:Ldepdisk}, MPDFs for case (a) are depicted. 
There are a few frames in which two or three peaks appear in an MPDF.
The largest peak and the second or third largest peak in the case of 
$(z_S,~R\sqrt{H_0/M_L}) = (0.4,~6)$, $(1.2,~2)$, and
$(2,~2)$ can be explained as the effect of the small lensing probability.
When the lensing probability is relatively small,
most of the light rays do not experience the lensing.
In this case, it is thus expected that the MPDF has a peak at 1.
This is because, 
in our definition of the magnification, when a 
light ray does not experience the lensing, 
the magnification is 1. 
When the light rays experience the lensing effect only once, 
the magnification 
typically becomes $\mu\sim R^4/(R^2-M_LD_S)^2$ as shown in 
Appendix \ref{sec:unimag}.
As a result, we find the second peak around $\mu\sim R^4/(R^2-M_LD_S)^2$,
which can be seen in Fig. \ref{fig:Ldepdisk}. 

In Figs. \ref{fig:Ldepdisk}-\ref{fig:Ldepsis}, 
we find that, in many cases, MPDFs can be
fitted using the gamma distribution $f(\mu-1;k,\theta)$ defined as
\begin{equation}
f(\mu-1;k,\theta)=(\mu-1)^{k-1}\frac{e^{-(\mu-1)/\theta}}{\theta^k\Gamma(k)}, 
\end{equation}
where $k>0$ and $\theta>0$ are called the shape parameter and scale parameter, 
respectively, and $\Gamma(k)$ is the gamma function.  
In these figures, we depict the gamma distributions that fit 
the MPDFs. 
The parameters of the gamma distributions are determined 
by the maximum likelihood method as follows.
We assume that the numerically generated magnifications are 
independent of each other 
and identically distributed random variables that obey 
the gamma distribution. 
The logarithm of the likelihood function $\mathcal L$ 
for $N$ samples 
of the magnification, $(\mu_1,\ldots,\mu_N)$ is given as
\begin{equation}
\ln \mathcal L(k,\theta)=\sum_{i=1}^N \ln f(\mu_i-1,k,\theta). 
\end{equation}

The shape parameter and the scale parameter are related 
to the mean $\mu_{\rm m}$ and variance $\sigma$ of the gamma distribution 
as follows:
\begin{eqnarray}
\mu_{\rm m}&=&k \theta+1, \\
\sigma^2&=& k \theta^2. 
\end{eqnarray}
The maximum likelihood estimates of the shape and scale parameters,
$\hat{k}$ and $\hat{\theta}$, are obtained by solving the equations,
\begin{eqnarray}
\frac{\partial}{\partial \theta}\ln \mathcal L(k,\theta)
\Big|_{k=\hat{k},\theta=\hat{\theta}}=0, 
\quad
{\frac{\partial}{\partial k}\ln \mathcal L(k,\theta)}
\Big|_{k=\hat{k},\theta=\hat{\theta}}=0.
\end{eqnarray}

\begin{figure}[htbp]
\begin{center}
\includegraphics[scale=1.3]{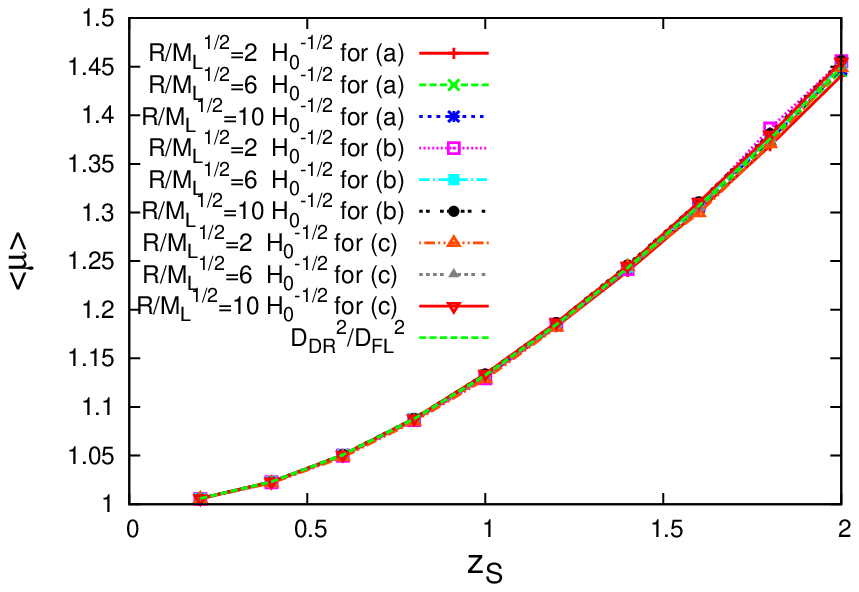}
\caption{The mean values of the magnification $\mu$ 
are depicted as functions of the source redshift.
}
\label{fig:mean}
\end{center}
\end{figure}

\begin{figure}[htbp]
\begin{center}
\includegraphics[scale=1.3]{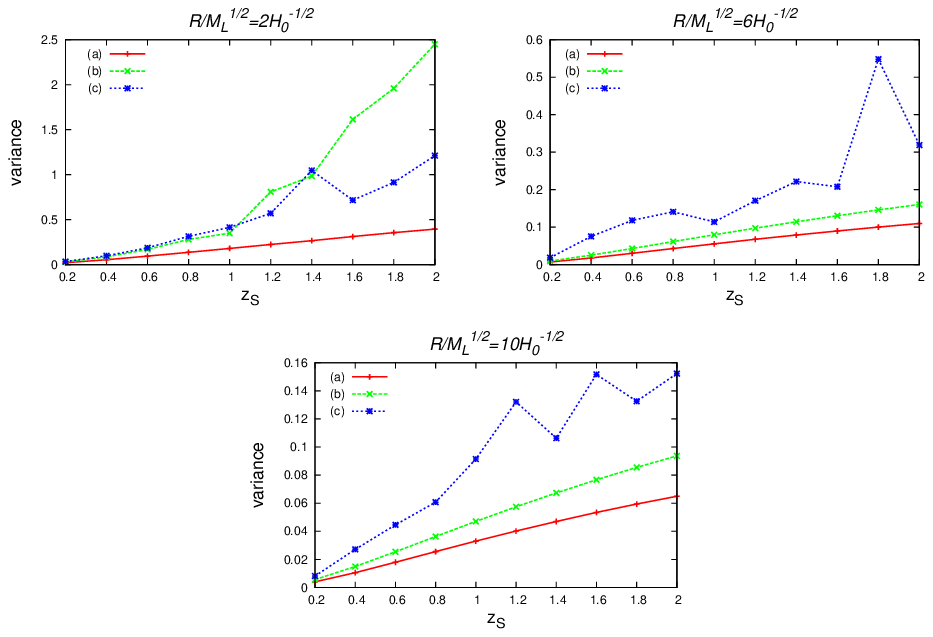}
\caption{The variances are depicted as functions of the source redshift.
}
\label{fig:variance}
\end{center}
\end{figure}

\begin{figure}[htbp]
\begin{center}
\includegraphics[scale=1.3]{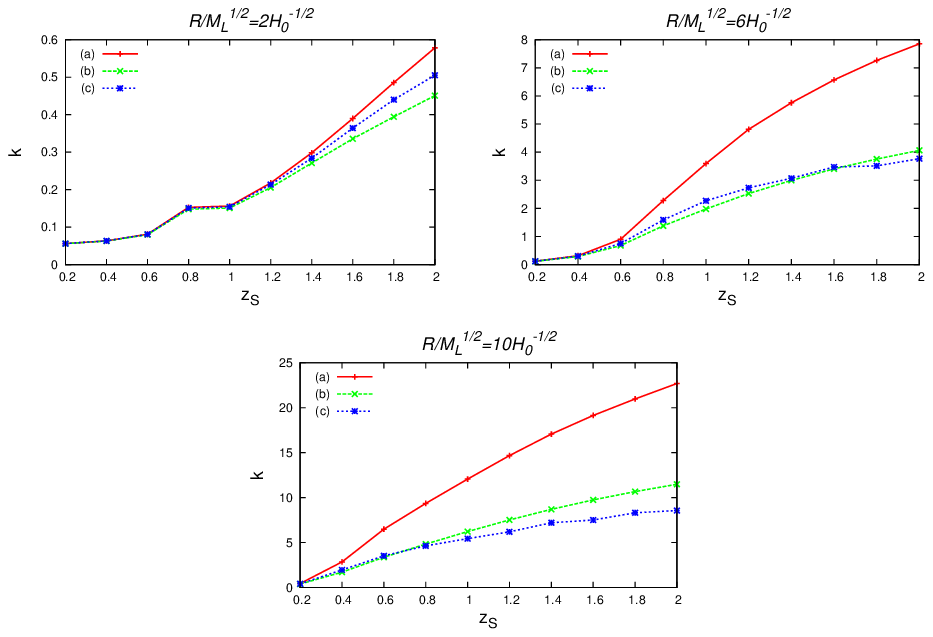}
\caption{The shape parameters $k$ are depicted as 
functions of the source redshift.
}
\label{fig:shape}
\end{center}
\end{figure}

\begin{figure}[htbp]
\begin{center}
\includegraphics[scale=1.3]{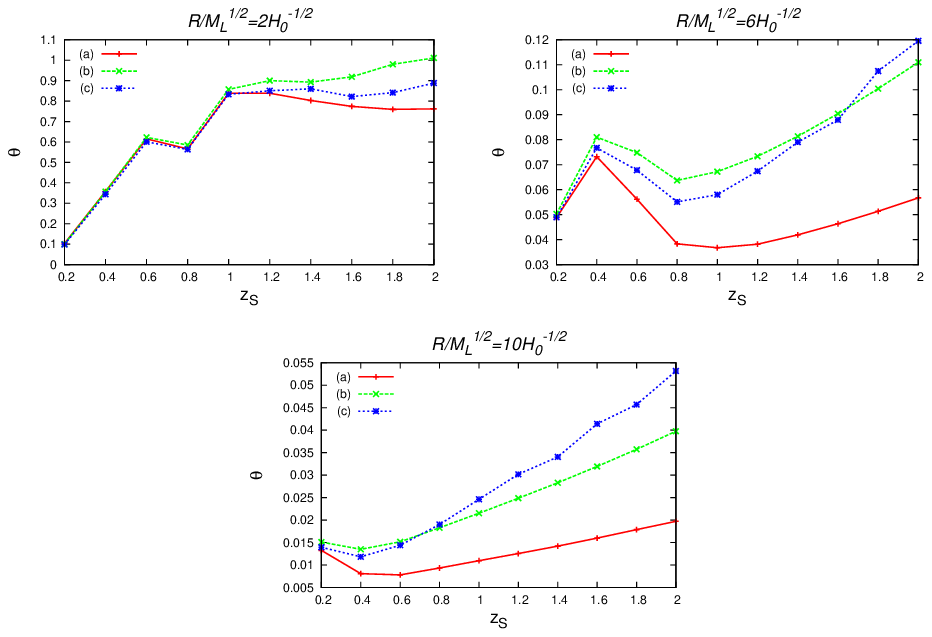}
\caption{The scale parameter $\theta$ is 
depicted as a function of the source redshift.
}
\label{fig:scale}
\end{center}
\end{figure}

The means $\left<\mu\right>$ and variances of the numerically generated MPDFs 
are plotted as functions of $z_S$ in Figs. \ref{fig:mean} 
and \ref{fig:variance}. 
The shape parameter $k$ and the scale parameter $\theta$ of 
the gamma distributions that fit the MPDFs
are also 
plotted as functions of $z_S$ in Figs. \ref{fig:shape} and \ref{fig:scale}. 
As explicitly shown in Fig. \ref{fig:mean}, 
$\left<\mu\right>$ 
is almost equal to 
$D_{DR}^2/D_{FL}^2$.  
The deviations are smaller than 1\% in all cases. 

In the case of $z_S=1.2$ and $z_S=2.0$ for the lens models (a) 
(see Fig. \ref{fig:Ldepdisk}),
the gamma distributions fit well the MPDFs for
$R/\sqrt{M_L}=6H_0^{-1/2},~10H_0^{-1/2}$
but do not for $R/\sqrt{M_L}=2H_0^{-1/2}$. 
In the case of model (b) and $z_S\geq 1.2$ (Fig. \ref{fig:Ldepnfw}), 
the gamma distribution fits well the MPDFs 
for $R/\sqrt{M_L}\geq 2H_0^{-1/2}$. 
In contrast, the gamma distributions do not fit
well the MPDFs in all cases for lens model (c)(see Fig. \ref{fig:Ldepsis}).
The lens model (c) has the steepest density profile at the center
among the lens models considered here.
The reason for the deviation from the gamma distributions
seems to come from this steep density profile.
This is consistent with the fact that, 
in the case of the point mass lens model, 
the gamma distributions do not fit the MPDFs. 
\footnote{In the case of point mass lens, 
the magnification probability behaves
$\sim \mu^{-3}$ when $\mu\gg 1$
(see, e.g., Ref.\citen{1984ApJ...284....1T} or \citen{1983ApJ...267..488V}). } 
This also suggests that, even in models (a) and (b), 
if we set $R/\sqrt{M_L}$ sufficiently smaller than 
$R/\sqrt{M_L}=2H_0^{-1/2}$, 
the gamma distributions do not fit the MPDFs irrespective of $z_S$.

Let us 
discuss the $z_S$ dependence of MPDFs. 
We consider the case of $R/\sqrt{M_L}=6H_0^{-1/2}$ 
for the lens model (a). 
While the gamma distribution fits well the MPDF at 
$z_S=2$, it does not at $z_S=0.4$. 
The most significant difference between these two cases is 
the number of lensing effects the light rays experience. 
Hence, our results suggest that, in order for a MPDF to be fitted well
by the gamma distribution, the light rays need to 
pass through near the clumps frequently. 
However, to clarify the condition for the realization of 
the gamma distribution more precisely, much more detailed investigations are 
required.

\begin{figure}[htbp]
\begin{center}
\includegraphics[scale=1.]{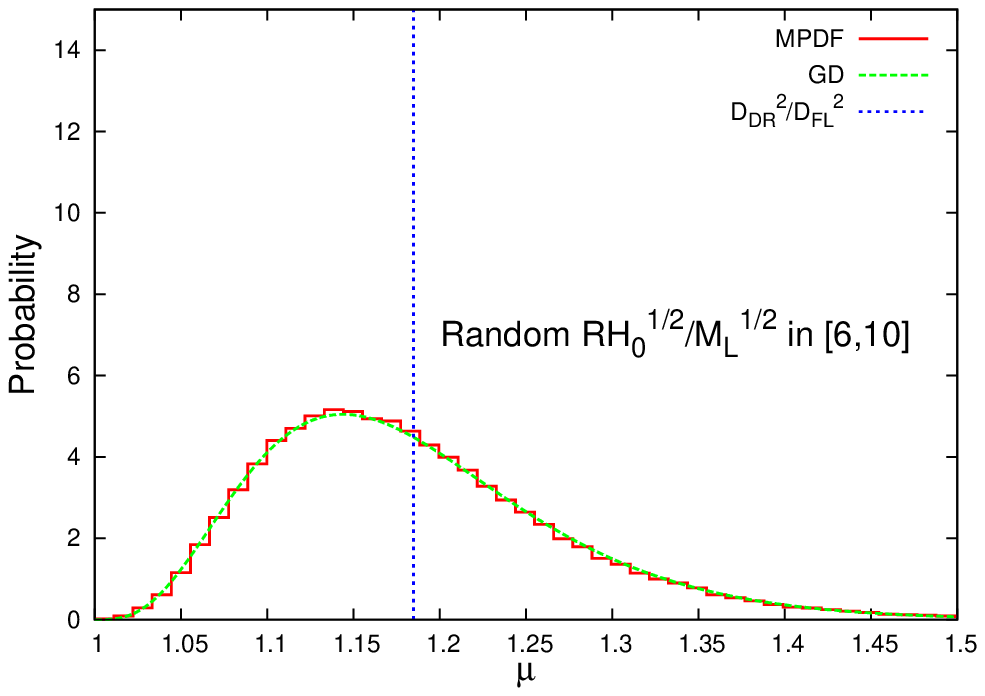}
\caption{ MPDF for the lens model (b) at 
$z_S=1.2$ is shown. 
Values of $R\sqrt{H_0/M_L}$ of clumps are distributed uniformly
within $6\leq R\sqrt{H_0/M_L}\leq 10$. 
The smooth lines are the gamma distributions that 
fit the MPDFs.
}
\label{fig:mix}
\end{center}
\end{figure}

The above results 
suggest that if lens models have sufficiently gradual density 
profiles, the MPDF is universally well fitted using the gamma distribution. 
To obtain further evidence for this hypothesis,
we calculate an MPDF in the case when 
the value of $R/\sqrt{M_L}$ is not constant. 
We assume the lens model (b) in which $R\sqrt{H_0/M_L}$ of 
each lens randomly takes a value within $6\leq R\sqrt{H_0/M_L}\leq10$, 
and that the redshift of the light source is given as $z_S=1.2$. 
The MPDF is shown in Fig. \ref{fig:mix}. 
One can find that MPDF is fitted well
by the gamma distribution even if 
the value of $R/\sqrt{M_L}$ is distributed. 
This fact might not be surprising, because 
the effect of changing the value of $R/\sqrt{M_L}$ is similar to the effect of 
changing the impact parameter of a light ray from a lens.
Both effects produce the change in the magnification. 
The shape parameter and the scale parameter become
$\theta=0.041$ and $k=4.5$, respectively. 
These are intermediate values between 
those in the cases of $R/\sqrt{M_L}=6H_0^{-1/2}$ 
and $R/\sqrt{M_L}=10H_0^{-1.2}$ for the lens model (b).

\subsection{$\chi^2$-test}\label{sec:chi2}
The discussions in \S\ref{sec:mpdfs} suggest that 
we may obtain the information about the slope of dark matter density profiles 
at the central cusp from the 
Type Ia supernovae observation. 
To clarify this possibility, we investigate the relation 
between the
goodness of fitting of the MPDFs to the gamma distributions 
and the steepness of the central density profile 
using the power law cusp lens model defined as
\begin{equation}
\Sigma(\xi)=\left\{
\begin{array}{ccc}
\frac{(2-n)M_L}{2\pi R^2}\left(\frac{R}{\xi}\right)^n&{\rm for}&\xi<R, \\
0&{\rm for}&\xi\geq R.  
\end{array}\right.\label{eq:pllens}
\end{equation}
This lens model includes the lens models (a) and (c) 
in \S\ref{sec:mpdfs} as the special cases. 
If we set $n=0$ or $1$ 
in the Eq. (\ref{eq:pllens}), this model becomes
model (a) or (c), respectively. 

We use the value of $\chi^2$ as an indicator of 
the goodness of fitting to the gamma distribution. 
Here, $\chi^2$ is defined as follows. 
The values of the magnification $\mu$ of each sample
are labelled with $i$ ($i=1,\ldots,N$) 
so that $\mu_i$ 
satisfies the relation, $\mu_1 \leq \mu_2 \leq \ldots \leq \mu_N$. 
We divide the region $\mu\geq 1$ into $p$ bins.
In this paper, we set $p=15$.  
The boundaries of the bins are expressed as $\nu_j$ ($j=0,\ldots,p-1$),
which satisfy the relation, $\nu_0< \nu_1< \ldots < \nu_{p-1}$,
where $\nu_0=1$.
We define $\nu_1=(\mu_m+\mu_{m+1})/2$ so that there are $m$ 
samples in the 1st bin.
In the same way, 
we define $\nu_{p-1}=(\mu_{N-m}+\mu_{N-m+1})/2$ so that there are $m$ 
samples in the $p$th bin. 
The values of $\nu_i$ from $i=2$ to $p-2$ are determined by 
even spacing. Namely, they are given as
\begin{eqnarray}
\nu_j&=&\nu_1+(j-1)\Delta\nu, \quad (j=2,\ldots,p-1) \\
\Delta\nu&=&(\nu_{p-1}-\nu_1)/(p-2).
\end{eqnarray}
The value of $m$ is determined 
such that each bin has a sufficiently large number of samples.  

\begin{figure}[htbp]
\begin{center}
\includegraphics[scale=1.]{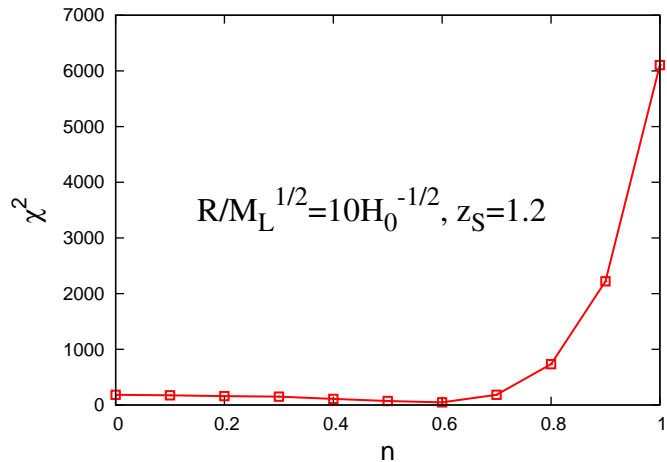}
\caption{The value of $\chi^2$ is depicted as 
the function of power $n$, where we set $R/\sqrt{M_L}=10H_0^{-1/2}$ 
and $z_S=1.2$. 
}
\label{fig:chi2}
\end{center}
\end{figure}

For $j$-th bin, we have the number of samples in this bin, 
$m_j$, and the expected number of samples, $e_j$, derived 
from the gamma distributions. 
We then define the value of $\chi^2$ as 
\begin{equation}
\chi^2=-2\sum^{15}_{i=1}m_i\log \frac{e_i}{m_i}. 
\end{equation}
In Fig. \ref{fig:chi2}, we depict the value of $\chi^2$ 
as the function of the steepness $n$ of the density 
profile, where $N=100,000$, and we have set 
$R/\sqrt{M_L}=10H_0^{-1/2}$, $z_S=1.2$, and $m=100$. 
The value of $\chi^2$ increases rapidly with $n$ if $n>0.6$. 

SNAP will be able to find the order of 
1000 SNe per year within the redshift $\sim 1$.
To simulate such situation, 
we divide the 100,000 samples into 100 datasets of 1000 samples,
and perform the $\chi^2$-test for each dataset. 
In this case, we set $m=10$. 
When the value of $\chi^2$ is larger than a certain threshold,
we judge that the MPDF cannot be fitted using the gamma distribution.
The threshold is determined such that there is only a 5 \% probability 
that the dataset generated from the gamma distribution is rejected. 
The relation between 
the rate of rejection and the value of $n$ is shown in Table \ref{tab:chi2}. 
We find that if $n\leq 0.7$, the rejection rate is around 10\% or less than that.
However, if $n\geq 0.8$, the rejection rate increases rapidly. 
This result suggests that we may constrain the slope of 
the dark matter density 
profile at the central cusp from the goodness of the fitting of the gamma distribution to the MPDF.

\begin{table}[htbp]
\begin{center}
\caption{The rate of rejection through the $\chi^2$-test 
with 1000 samples is shown in this table. 
}
\label{tab:chi2}
\begin{tabular}{|c|c|c|c|c|c|c|c|c|c|c|c|}
\hline
Power $n$&0&0.1&0.2&0.3&0.4&0.5&0.6&0.7&0.8&0.9&1.0\\
\hline
Rate of rejection&$0.1$&$0.06$&$0.06
$&$0.05$&$0.12$&$0.14$&$0.07
$&$0.11
$&$0.27$&$0.82$&$0.97$\\
\hline
\end{tabular}
\end{center}
\end{table}

\section{Summary and conclusions}\label{sec:consum}
We have studied lensing effects on observations of point sources 
such as Type Ia supernovae by the Monte-Carlo simulation. 
It has been assumed that all matter in the universe 
takes the form of randomly distributed objects each of 
which has finite size and transparent to light rays. 
In addition, we have assumed that each of the lens objects is 
axially symmetric along the line of sight.
We have found analytically that 
the magnification probability distribution functions (MPDFs)
depend on the mass $M_L$ and the size $R$ 
of a lens object only through 
the form $R/\sqrt{M_L}$.
We have calculated MPDFs 
for various values of the source redshift  and 
$R/\sqrt{M_L}$. 
We found that the resultant MPDFs can be categorized
into two groups according to 
whether the gamma distributions fit well the MPDFs or not. 
In the case of the lens models that have smooth 
density profiles, the gamma distributions fit well 
the MPDFs 
if $R\sqrt{H_0/M_L}$ is sufficiently large. 
Furthermore, we have found that this result holds in the lens model 
that has the same degree of density concentration 
at the center as the Navarro-Frenk-White (NFW) lens model. 
In contrast, the gamma distributions do not fit well 
the MPDFs for 
any value of $R/\sqrt{M_L}$ 
in the case of the singular isothermal sphere (SIS) lens model. 
These results suggest that we might be able to distinguish 
the NFW lens model from the SIS lens model on the basis of MPDFs.

We have shown that MPDF is well fitted using the gamma distribution 
even if the value of $R/\sqrt{M_L}$ is uniformly distributed 
within $6\leq R/\sqrt{M_L}\leq 10$ in the case of 
lens model (b). 
This result suggests that MPDF is universally well fitted 
using the gamma distribution if lens models 
have sufficiently gradual density profiles. 
To obtain further evidence for this hypothesis, 
we need to clarify the effect of the distribution of $R/\sqrt{M_L}$ 
in more detail. 
This is one of our future works.

We introduced the power law cusp lens model, 
and we investigated the dependence of 
the goodness of the fitting with the gamma distribution 
on the power $n$, which represents the steepness of the cusp at the center. 
We have generated 100,000 samples of magnification 
in the Monte-Carlo simulations. 
We divided the 100,000 samples into 100 datasets of 1000 samples,
and performed the $\chi^2$-test for each dataset,
since the same analysis might be possible for the observational 
data of SNAP. 
Then, we found the significant difference in 
the rate of rejection between 
$n\leq 0.7$ and $n\geq0.8$. 
This result suggests that we may obtain the information 
about a slope of the density profiles of the central cusp of 
dark matter using 
MPDF of Type Ia supernovae. 

A mathematical explanation of the reason why we obtain the gamma distribution 
is unavailable now. 
In Refs.\citen{Wang:2002qc,Bergstrom:1999xh,2007arXiv0711.0743L}, and 
\citen{Wang:2004ax}, 
analytic fitting functions different from ours for MPDF are proposed. 
The relation between our analytic fitting function and 
those should be studied. 
We note that 
we need to examine the effect of simplifications carried out in this study,
before we compare our results with observations. 
For example, we have assumed that there are no spatial correlations 
of the distribution of the clumps. 
The effect of the spatial correlation of the clumps should be clarified. 
We leave these issues as our future works. 

The density profile of dark matter halos is a hot topic in 
astrophysics both theoretically and observationally 
(See e.g. Ref.\citen{BinneyTremaine}).
Our results suggest that, 
although there are some simplifications in 
comparison with realistic situations of our universe,
the statistical gravitational lensing effect
may shed a new light on the observational investigation of the density 
profile of dark matter halos.

\section*{Acknowledgements}
We would like to acknowledge the helpful advice of Professor M. Sasaki and 
Dr. H. Kozaki. 
C. Yoo was supported by the 21st COE program ``Constitution of wide-angle mathematical basis focused on knots" from the Japan Ministry of Education, 
Culture, Sports, Science and Technology. 
H.T's work was supported in part by JSPS, KAKENHI, Nos.16540251 and 20540271. 
This work was supported in part by a JSPS Grant-in-Aid
for Scientific Research (B), No.~17340075.

\appendix

\section{Calculation Method}\label{sec:cal}
To calculate magnification factors,
we use the multiple lens-\hspace{0pt}plane method~\cite{SEF}.
We consider the ``straight'' line $A$ from the source to the observer, 
and we put $N$ lens planes between them so that the straight line $A$
intersects vertically.
We denote lens planes from the observer to the source
sequentially as $\Sigma_1, \Sigma_2, \dots, \Sigma_N$.
For convenience, we label the source plane that is also orthogonal to
the line $A$ as $\Sigma_{N + 1}$.
On each lens plane, say $\Sigma_i$, the lens position $\bzeta_i$ and
the ray position $\bgamma_i$ are specified with respect to the
intersection point of $A$.

Suppose lens positions $\bzeta_i~(i = 1, \dots, N)$ are given.
When we emanate a ray with a ray position $\bgamma_1$ on $\Sigma_1$,
the lens position on the $j$-th lens plane $\Sigma_j~~(j \leq N + 1)$
is recursively given as follows
\begin{equation}
 \displaystyle
\bgamma_j
 =   \frac{D_j}{D_1} \bgamma_1
   - \sum^{j-1}_{i=1} D_{ij} \hat{\balpha}(\bgamma_i - \bzeta_i),
 \label{eq:tikuji} 
\end{equation}
where $D_{ij}$ is the DR distance from the $i(<j)$-th
plane to the $j$-th plane.
The right hand side of equation~(\ref{eq:tikuji}) also includes ray positions
$\bgamma_2, \dots, \bgamma_{j-1} $.
Thus, we must calculate the ray positions
$\bgamma_2, \dots, \bgamma_{j - 1}$ to obtain the ray position
$\bgamma_{j}$.
For later convenience,
we specify the ray positions and lens positions using the
following dimensionless quantities:
\begin{eqnarray}
\displaystyle
 \bu_i &:=& \frac{\bgamma_i}{D_i}, \label{eq:u-def}\\
 \bq_i &:=& \frac{\bzeta_i}{D_i}.
\end{eqnarray}
Equation~(\ref{eq:tikuji}) is written as
\begin{equation}
 \bu_j
 =   \bu_1
   - \sum^{j-1}_{i=1} \beta_{ij} \balpha(\bu_i - \bq_i),
   \label{eq:tikuji2}
\end{equation}
where 
\begin{equation}
\displaystyle
 \beta_{ij} := \frac{D_{ij} D_S}{D_j D_{iS}}
\end{equation}
and
\begin{equation}
 \displaystyle
\balpha(\bu_i - \bq_i)
 := \frac{D_{iS}}{D_S} \hat{\balpha}(\bgamma_i - \bzeta_i).
\end{equation}

\begin{figure}[htbp]
\begin{center}
\includegraphics[scale=0.4]{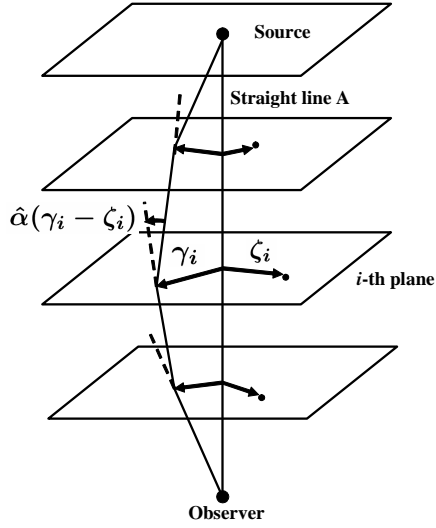}
\caption{The definitions of vectors $\balpha_j$, $\bzeta_j$, and
 $\bgamma_j$ are given here. 
}
\label{fig:multi}
\end{center}
\end{figure}

In the case of the multiple lensing effect,
multiple images appear in general.
We specify the paths from the source to the observer using 
their
ray positions, say $\bov_{(p)}$, on the first lens plane.
For a path $\bov_{(p)}$, the magnification $\mu_{(p)}$ is
defined as
\begin{equation}
\displaystyle
 \mu_{(p)} = \frac{1}{\det \mathcal{A}_{(p)}},
 \label{eq:mup}
\end{equation}
where $\mathcal{A}_{(p)}$ is the Jacobian matrix of the lens mapping:
\begin{eqnarray}
\displaystyle
 \mathcal{A}_{(p)}
 &=& \left.
      \frac{\partial \bu_{N + 1}}{\partial \bu_{1}}
    \right|_{\bu_1 = \bov_{(p)}} \nonumber \\
 &=& \mathcal{I}
      - \sum^{N}_{i=1} \beta_{i, N+1}
          \left.
            \frac{\partial \balpha(\bu_i - \bq_i)}{\partial \bu_i}
            \frac{\partial \bu_i}{\partial \bu_1}
          \right|_{\bu_1 = \bov_{(p)}}.
 \label{eq:jaco}
\end{eqnarray}

We obtain the following result. 

\vskip0.3cm
\noindent
Proposition. {\it The magnification probability distribution function 
for the identical lens objects 
is invariant under the following constant rescaling of the linear 
extent $R$ and the mass $M_L$ of the lens object  
\begin{equation}
R\rightarrow CR ~~{\rm and}~~ M_L\rightarrow C^2M_L,  \label{eq:rescaling}
\end{equation}
if the surface mass densities of the lens objects are given in the form of 
Eq.(\ref{eq:lensform}). }

\vskip0.3cm
{\it Proof.} The parameter $\xi_0$ becomes $C\xi_0$ through the changes of the parameters 
(\ref{eq:rescaling}) due to its definition (\ref{eq:re}). 
Hence, we find from Eq. (\ref{eq:p3}) that the density 
$p(y,z)$ of lenses in $(y,z)$ space is unchanged through the 
replacement (\ref{eq:rescaling}), if 
$\zeta$ is also replaced by $C\zeta$; hereafter we assume so. 

Substituting Eq. (\ref{eq:lensform}) into Eq. (\ref{eq:defangle}), we have
\begin{equation}
\hat{\bolm\alpha}(\bolm\xi)
={\rm Const}\times \frac{4M_L}{R^2}\int_{\xi'\leq R}\frac{(\bolm\xi-\bolm\xi')
F(R/\xi')}{|\bolm\xi-\bolm\xi'|^2}d^2\xi'. 
\end{equation}
For the changes in the parameters (\ref{eq:rescaling}), $\hat{\bolm\alpha}$ changes 
in the manner 
\begin{eqnarray}
\hat{\bolm\alpha}(\bolm\xi)
&&\longrightarrow 
{\rm Const}\times \frac{4M_L}{R^2}\int_{\xi'\leq CR}\frac{(\bolm\xi-\bolm\xi')
F(CR/\xi')}{|\bolm\xi-\bolm\xi'|^2}d^2\xi' \nonumber \\
&&={\rm Const}\times \frac{4M_L}{R^2}\int_{\bar{\xi}\leq R}\frac{(\bolm\xi-C\bar{\bolm\xi})
F(R/\bar{\xi})}{|\bolm\xi-C\bar{\bolm\xi}|^2}C^2d^2\bar{\xi} \nonumber \\
&&=C\hat{\bolm\alpha}(C^{-1}\bolm\xi), \label{eq:rescaled-alpha}
\end{eqnarray}
where in the second line, we have replaced the variable of integration $\bolm\xi'$ 
by $\bar{\bolm\xi}=C^{-1}\bolm\xi'$. 

The solution of Eq. (\ref{eq:tikuji}) with the parameters changed as 
Eq. (\ref{eq:rescaling}) is 
denoted by $\tilde{\bolm\gamma}_j$. 
By using Eq. (\ref{eq:rescaled-alpha}), 
the equations satisfied by $\tilde{\bolm\gamma}_j$ are 
given as 
\begin{equation}
 \displaystyle
\tilde{\bolm\gamma}_j
 =   \frac{D_j}{D_1} \tilde{\bolm\gamma}_1
   - \sum^{j-1}_{i=1} D_{ij}C \hat{\balpha}
\left(C^{-1}\left[\tilde{\bolm\gamma}_i - C\bzeta_i\right]\right). 
\end{equation}
It is easily confirmed that $\tilde{\bolm\gamma}_j=C\bolm\gamma_j$
is the solution for the above equation, where $\bolm\gamma_j$ 
is the solution of 
Eq. (\ref{eq:tikuji}). Thus, for the changes in the parameters of lens objects 
(\ref{eq:rescaling}), we have the following scaling law
\begin{equation}
\bu_j\longrightarrow C\bu_j, \label{eq:rescaled-u} 
\end{equation}
 (see Eq. (\ref{eq:u-def})). 
Since the Jacobian matrix $\mathcal{A}_{(p)}$ is invariant for the constant rescaling  
(\ref{eq:rescaled-u})  due to definition (\ref{eq:jaco}), 
we eventually find that the Jacobian matrix $\mathcal{A}_{(p)}$
does not change through the replacement (\ref{eq:rescaling}). 
This result means that the magnification $\mu_{(p)}$ is invariant 
for the change in parameters in the form of 
 (\ref{eq:rescaling}) and thus the magnification probability distribution 
function is also invariant for (\ref{eq:rescaling}). \hfill $\Box$

\vskip0.3cm
The above result guarantees that the MPDF 
depends on only the parameter $R/\sqrt{M_L}$. 

We randomly put lenses so that the distribution
of those is consistent with equation~(\ref{eq:p3}).
First, we divide the spherical region $z<z_S$ into $N$ concentric
spherical shells each of which is bounded by two spheres
$z = z_i - \Delta z / 2$ and $z = z_i + \Delta z / 2$.
We take into account only the nearest lens in each shell.
We find from equation~(\ref{eq:p3}) that in the $i$-th shell
$z_i - \Delta z / 2 < z < z_i + \Delta z / 2$ there is one point mass on
average within the region $y\leq Y_i$, where 
\begin{equation}
 \int^{Y_i}_0 p(y, z_i) dy=1.
\end{equation}
Therefore 
we randomly put a point mass within the region $y\leq Y_i$ in the
$i$-th shell.
We can neglect the lensing effects of the clumps that are so far
from the ray that they do not affect the magnification.
We set an upper bound $y_{\rm max}~~ (< Y_{i})$ to the lens position,
and take account of lensing effects due to the lenses
in the region $y < y_{\rm max}$.
We have set the value of $y_{\rm max}$ as 
\footnote{
We have determined the value of $y_{\rm max}$ 
as sufficiently large 
so that $<\mu>\simeq D^2_{DR}/D^2_{FL}$ 
in the case of the power law tail lens model. }
\begin{equation}
y_{\rm max}=\max\left\{5,\frac{R}{\xi_0}+1\right\}. \label{eq:ymax}
\end{equation}

Next, we show the method to find paths from the source to the
observer: $\bov_{(p)}$.
When lens positions are given, 
the dimensionless ray position $\bu_{N+1}$ on the source plane
$\Sigma_{N+1}$ is given as a function of only the ray position $\bu_{1}$ on the
first lens plane $\Sigma_{1}$;
\begin{equation}
 \bu_{N+1} = \bolm{f}(\bu_{1}). \label{eq:lens mapping}
\end{equation}
Therefore, to find $\bov_{(p)}$, we solve the equation:
\begin{equation}
 \bu_{N+1}(\bu_{1}) = 0. \label{eq:map0}
\end{equation}
We find the roots of equation~(\ref{eq:map0}) by 
the Newton-\hspace{0pt}Raphson method. The technical details can be seen in 
Ref.\citen{1991ApJ...374...83R}.

\section{Typical Value of the Magnification 
for the Lens Model (a)}\label{sec:unimag}
In the case of the lens model (a), 
from Eq. (\ref{eq:defangle}), 
the bending angle vector $\hat \balpha(\bxi)$ is given as 
\begin{equation}
\hat \balpha(\bxi)=\frac{4M_L}{R^2}\bxi. 
\end{equation}
Since the magnification $\mu$ due to only one lens plane 
is given as 
\begin{equation}
\mu=\left|\det \left(\frac{\partial \boeta}{\partial \bxi}\right)\right|^{-1}
\frac{D_S^2}{D_L^2}, 
\end{equation}
we have 
\begin{equation}
\mu=\frac{R^4}{\left(R^2-\frac{4M_LD_LD_{LS}}{D_S}\right)^2}
\end{equation}
for the lens model (a) from Eq. (\ref{eq:lens}). 
Using $D_S$ instead of $2D_L$ or $2D_{LS}$, 
we have 
\begin{equation}
\mu\sim\frac{R^4}{\left(R^2-M_LD_S\right)^2}. 
\end{equation}


\begin{thebibliography}{}
\bibitem{Perlmutter:1998np} 
  S. Perlmutter et al., \AJ{517,1999,565}. 
\bibitem{Riess:1998cb} 
  A.~G. Riess et al., \AJ{116,1998,1009}.  
\bibitem{Knop}
  R.~A.~Knop, et al., 
  \AJ{598,2003,102}. 
\bibitem{Riess:2004nr}
  A.~G.~Riess et al., 
  \AJ{607,2004,665}. 
\bibitem{Sarkar:2007sp}
  D.~Sarkar, A.~Amblard, D.~E.~Holz and A.~Cooray,
  arXiv:0710.4143.
\bibitem{Holz:2004xx}
  D.~E.~Holz and E.~V.~Linder,
  \AJ{631,2005,678}. 
\bibitem{1991ApJ...374...83R} 
  K.~P. Rauch, \AJ{374,1991,83}. 
\bibitem{1997ApJ...475L..81W} 
  J. Wambsganss, R. Cen, 
  G. Xu and J.~P. Ostriker, \AJ{475,1997,L81}. 
\bibitem{2006PhRvL..96b1301C} A.~Cooray, D.~Huterer 
and D.~E.~Holz, \PRL{96,2006,021301}. 
\bibitem{2000MNRAS.318..195B} 
  A.~J.~Barber, 
  Mon.\ Not.\ R.\ Astron.\ Soc.\ {\bf 318} (2000), 195.
\bibitem{2000ApJ...532..679P} 
  C.~Porciani and P.~Madau, \AJ{532,2000,679}. 
\bibitem{2000A&A...354..767V} 
  P.~Valageas, Astron. Astrophys. {\bf 354} (2000), 767. 
\bibitem{1998ApJ...506L...1H} 
  D.~E.~Holz, \AJ{506,1998,L1}. 
\bibitem{Frieman:1996xk}
  J.~A. Frieman, Comments Astrophys.
  \textbf{18} (1996), 323. 
\bibitem{Martel:2007fh}
  H.~Martel and P.~Premadi, 
  \AJ{673,2007,657}. 
\bibitem{1999ApJ...519L...1M} R.~B. Metcalf and 
  J. Silk, \AJ{519,1999,L1}. 
\bibitem{2001ApJ...559...53M} 
  E. M{\"o}rtsell, 
  A. Goobar and L. Bergstr{\"o}m, \AJ{559,2001,53}.   
\bibitem{1999A&A...351L..10S} 
  U. Seljak  and  D.~E. Holz, 
 Astron. Astrophys. {\bf 351} (1999), L10. 
\bibitem{2001ApJ...556L..71H} 
  D.~E.~Holz, \AJ{556,2001,L71}. 
\bibitem{2000ApJ...534...29H} 
  T.~Hamana and T.~Futamase, \AJ{534,2000,29}. 
\bibitem{1999MNRAS.305..746M} 
  R.~B.~Metcalf, 
  Mon.\ Not.\ R.\ Astron.\ Soc.\ {\bf 305} (1999), 746. 
\bibitem{2007arXiv0711.0743L} E.~V. Linder, arXiv:0711.0743. 
\bibitem{2007JCAP...06....2J} 
  J.~J{\"o}nsson, T.~Dahl{\'e}n, A.~Goobar, E.~M{\"o}rtsell 
  and A.~Riess, J. Cosmol. Astropart. Phys.
   {\bf 06} (2007), 002. 
\bibitem{2005MNRAS.358..101M} 
  B.~M{\'e}nard and N.~Dalal, 
  Mon.\ Not.\ R.\ Astron.\ Soc.\  {\bf 358} (2005), 101.
\bibitem{Wang:2004ax}
  Y.~Wang,
  J. Cosmol. Astropart. Phys.
 {\bf 03} (2005), 005.
\bibitem{2004MNRAS.351.1387W} 
  L.~L.~R.~Williams and J.~Song, 
  Mon.\ Not.\ R.\ Astron.\ Soc.\ {\bf 351} (2004), 1387.
\bibitem{1998PhRvD..58f3501H} 
  D.~E. Holz and R.~M. Wald, 
  \PRD{58,1998,063501}. 
\bibitem{2007PhRvL..98g1302M} 
  R.~B. Metcalf and 
  J. Silk \PRL{98,2007,071302}. 
\bibitem{BinneyTremaine} 
  J. Binney and S. Tremaine, 2008 
{\em Galactic Dynamics, 2nd Edition}
 (Princeton University Press),  p. 751. 
\bibitem{2006astro.ph..1683M} 
  D. Munshi and P. Valageas, 
  astro-ph/0601683. 
\bibitem{2002MNRAS.335.1061S} 
  M.~Sereno, E.~Piedipalumbo and M.~V.~Sazhin, 
  Mon.\ Not.\ R.\ Astron.\ Soc.\  {\bf 335} (2002), 1061. 
\bibitem{2000PhLB..486..249G} 
  M.~Goliath and E.~{\ M\"o}rtsell, 
  \PLB{486,2000,249}. 
\bibitem{Wang:2002qc} Y. Wang, D.~E. Holz and 
  D. Munshi, \AJ{572,2002,L15}. 
\bibitem{Gunnarsson:2003fy}
  C.~Gunnarsson, J. Cosmol. Astropart. Phys. \textbf{03} (2004), 002. 
\bibitem{Wang:1999bz}
  Y.~Wang, \AJ{536,2000,531}.
\bibitem{Wang:2003gz}
  Y.~Wang and P.~Mukherjee, \AJ{606,2004,654}. 
\bibitem{SEF}
  P. Schneider, J. Elers and E. E. Falco,  1992
  {\em Gravitational Lenses}~
  (New York: Springer, 1992)
\bibitem{Yoo:2006gn} 
  C.-M. Yoo, K. Nakao, H. Kozaki and R. Takahashi, \AJ{655,2007,691}. 
\bibitem{NFW}
  J. F. Navarro, C. S. Frenk and S. D. M. White, \AJ{490,1997,493}. 
\bibitem{1973ApJ...180L..31D} 
  C.~C. Dyer and R.~C. Roeder, \AJ{180,1973,L31}. 
\bibitem{1969ApJ...155...89K} 
  R. Kantowski, \AJ{155,1969,89}. 
\bibitem{Jain}
  B.~Jain, U.~Seljak and S.~White, 
  \AJ{530,2000,547}. 
\bibitem{Wambsganss} 
  J.~Wambsganss, R.~Cen and 
  J.~P.~Ostriker, \AJ{494,1998,29}. 
\bibitem{Premadi}
  P.~Premadi, H.~Martel and R.~Matzner, 
  \AJ{493,1998,10}.
\bibitem{Fukugita}
  M.~Fukugita, T.~Futamase, M.~Kasai and E.~L.~Turner, \AJ{393,1992,3}. 
\bibitem{Asada}
  H.~Asada, \AJ{501,1998,473}. 
\bibitem{Kasai}
  M.~Kasai, T.~Futamase and F.~Takahara, \PLA{147,1990,97}. 
\bibitem{Watanabe}
  K.~Watanabe and K.~Tomita, \AJ{355,1990,1}.
\bibitem{Tomita}
  K.~Tomita, \PTP{100,1998,79}. 
\bibitem{TAH}
  K.~Tomita, H.~Asada and T.~Hamana, 
  \PTPS{133,1999,155}. 
\bibitem{Seitz1994}
  S.~Seitz, P.~Schneider and J.~Ehlers, 
  Class. Quantum Grav. {\bf 11} (1994), 2345.
\bibitem{Kozaki}
  H. Kozaki, private communication.
\bibitem{1984ApJ...284....1T} 
  E.~L. Turner, J.~P. Ostriker and J.~R. Gott, III, \AJ{284,1984,1}. 
\bibitem{1983ApJ...267..488V} 
  M. Vietri and J.~P. Ostriker, \AJ{267,1983,488}. 
\bibitem{Bergstrom:1999xh} 
  L. Bergstr{\"o}m, M. Goliath, A. Goobar and E. M{\"o}rtsell, 
  Astron. Astrophys. \textbf{358} (2000), 13.
\end{thebibliography}
\end{document}